%% file: article.tex
\documentclass[12pt]{article}
\usepackage{epsfig}
\usepackage{amsmath}
\usepackage{graphicx}
\usepackage{url}
\usepackage{hyperref}
\RequirePackage{xspace}
\usepackage{tabularx}

\input econfmacros.tex
\def\invfb{\ensuremath{\mbox{\,fb}^{-1}}\xspace}
\def\Y#1S{\ensuremath{\Upsilon{(#1S)}}\xspace}
\def\Title#1{\begin{center} {\Large {\bf #1} } \end{center}}

\begin{document}

\Title{Recent Results on Radiative and Electroweak Penguin decays at Belle}

\bigskip\bigskip

\begin{center}
{\it S. Sandilya\index{Sandilya, S.}\\
(On behalf of the Belle Collaboration)\\
University of Cincinnati,\\
Cincinnati, Ohio 45221}
\end{center}
\bigskip\bigskip

\begin{abstract}
  We report on the measurement of the decay $B\to K^{\star}\gamma$, lepton-flavor dependent angular analysis of the decay $B\to K^{\star}\ell\ell$, and search for the decays $B\to h\nu\nu$. All these analyses are performed on the 711 \invfb data sample recorded by the Belle detector at the \Y4S resonance. 
\end{abstract}

\section{Introduction}
\label{sec:intro}
Flavor changing neutral current (FCNC) $B$ decays are forbidden at tree level in the standard model (SM) and proceed at lowest order through penguin loop and box diagrams.
These types of rare decays are sensitive to potential contributions from non-SM particles that can enter into the loop diagram.
The Belle experiment recently published notable results on the measurements of the decays involving radiative and electroweak penguin $B$ meson decays mediated by the $b \to s$ transition.
All these studies are based on the full data set recorded at the \Y4S resonance by the Belle detector at the KEKB energy-asymmetric collider; this sample of 711 \invfb luminosity contains $772 \times 10^{6}$ \BB pairs.
We report herein the recent measurements of the decay $B\to K^{\star}\gamma$ ~\cite{btosg:belle}, lepton-flavor dependent angular analysis of the decay $B\to K^{\star}\ell\ell$~\cite{btosll:belle}, and search for the decays $B\to h\nu\nu$~\cite{btohnn:belle}.

\section{Measurement of $B\to K^{\star}\gamma$}
\label{sec:btosg}
The decays $B\to K^{\star}\gamma$ [$K^{\star}$ refers to $K^{\star}(892)$] involves the quark-level transition $b\to s \gamma$ and proceed dominantly via one-loop electromagnetic penguin diagrams.
Among the decays involving such transitions, $B\to K^{\star}\gamma$ is the cleanest one as the $K^{\star}$ resonance mass is well separated from the other higher $K\pi$ resonances and the decay has a large branching fraction (BF).
The SM prediction for BF of these decays has large uncertainity ($\sim 30\%$) due to form factor and gives a weak constraint on new physics (NP)~\cite{{btosg:th1},{btosg:th2},{btosg:th3},{btosg:th4},{btosg:th5},{btosg:th6},{btosg:th7}}.
However, these theory uncertainties cancel out in the BF ratios, like isospin ($\Delta_{0+}$) and direct $CP$ asymmetries ($A_{CP}$), and provide strong constraint on NP ~\cite{btosg:th5}.
The $\Delta_{0+}$, $A_{CP}$ and the difference of $A_{CP}$ between the charged and neutral $B$ mesons ($\Delta A_{CP}$) are defined as follows:

\begin{equation}
  \Delta_{0+} = \frac{\Gamma(B^{0}\to K^{\star 0}\gamma) - \Gamma(B^{+}\to K^{\star +}\gamma)}
        {\Gamma(B^{0}\to K^{\star 0}\gamma) + \Gamma(B^{+}\to K^{\star +}\gamma)}
        \label{eq:1}
\end{equation}

\begin{equation}
  A_{CP} = \frac{\Gamma(\Bb\to \Kb^{\star}\gamma) - \Gamma(B\to K^{\star}\gamma)}
  {\Gamma(\Bb\to \Kb^{\star}\gamma) + \Gamma(B\to K^{\star}\gamma)}
  \label{eq:2}
\end{equation}

\begin{equation}
  \Delta A_{CP} = A_{CP}(B^{+}\to K^{\star +}\gamma) - A_{CP}(B^{0}\to K^{\star 0}\gamma)
  \label{eq:2}
\end{equation}
where $\Gamma$ denotes the partial width of the corresponding decay.

\begin{figure}[htb]
\includegraphics[width=0.3\textwidth]{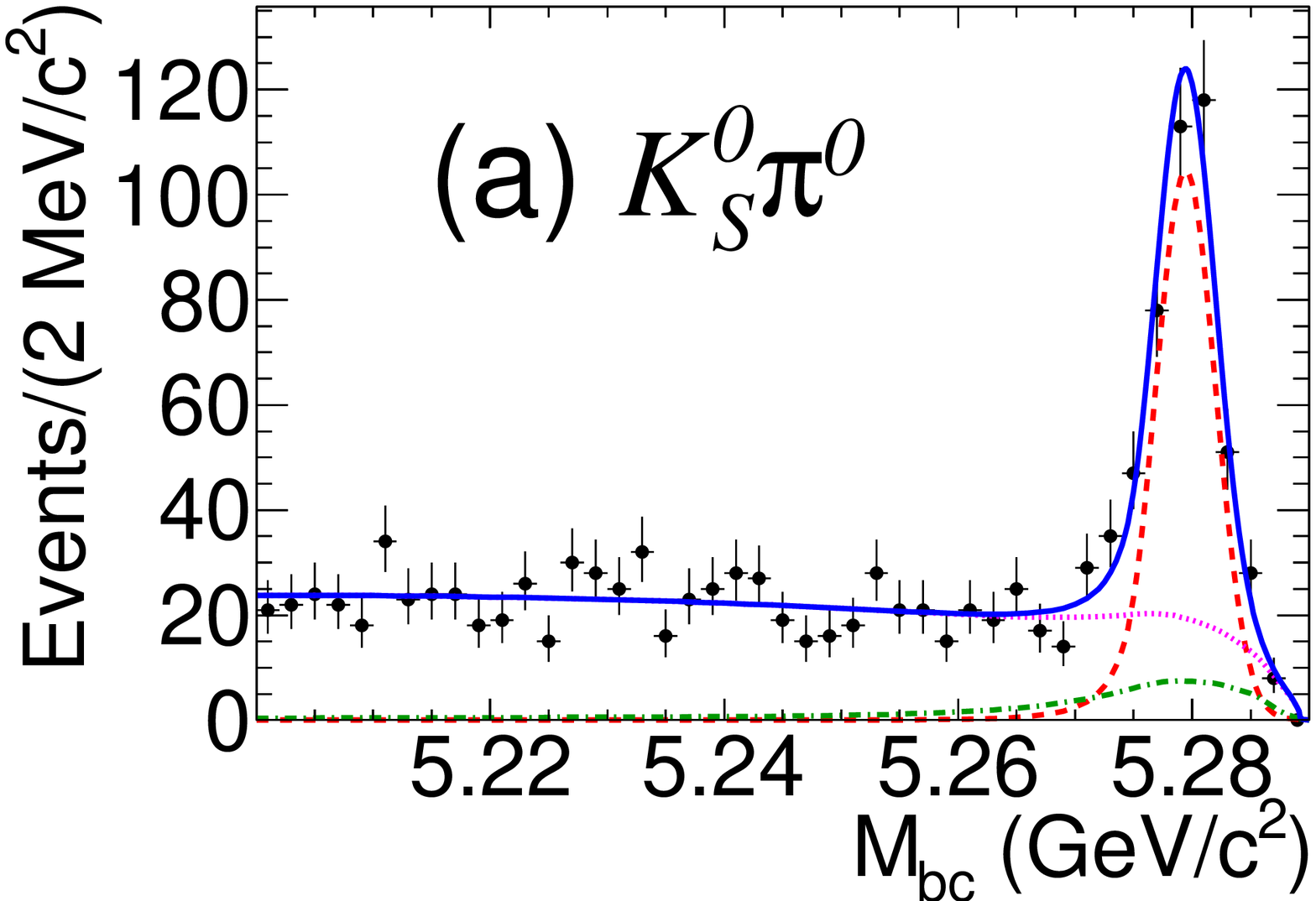}\\ 
\includegraphics[width=0.3\textwidth]{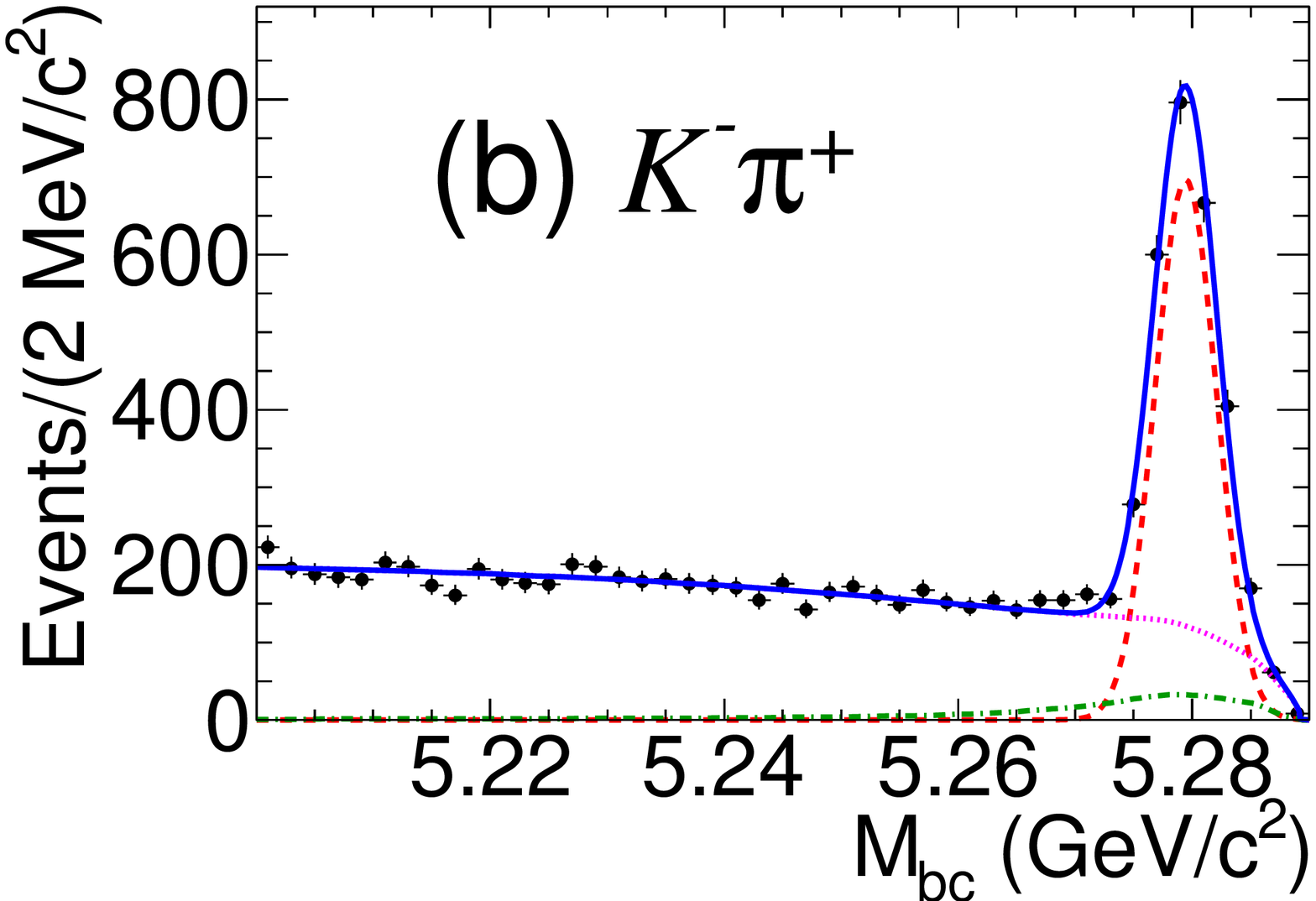}   
\includegraphics[width=0.3\textwidth]{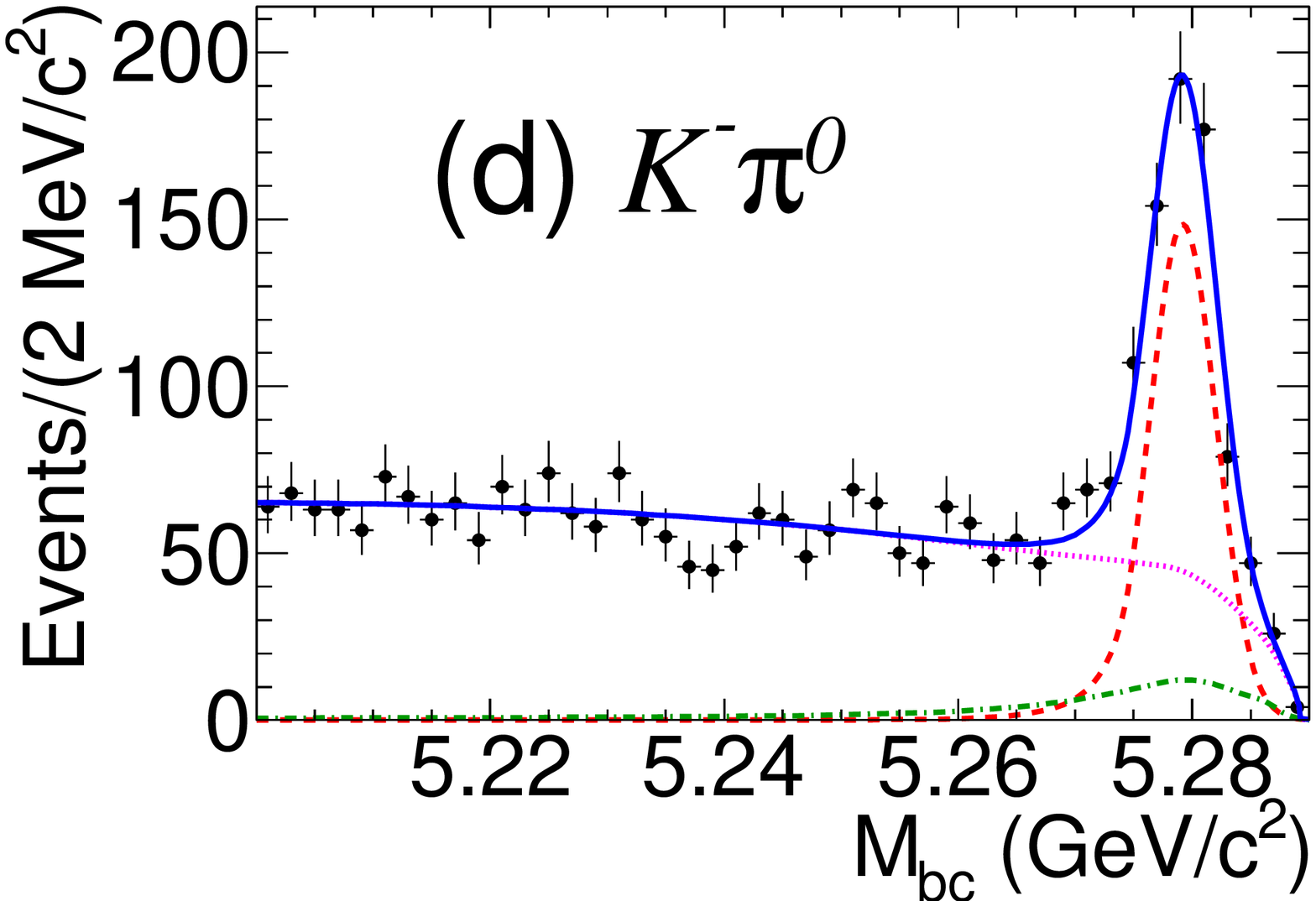}   
\includegraphics[width=0.3\textwidth]{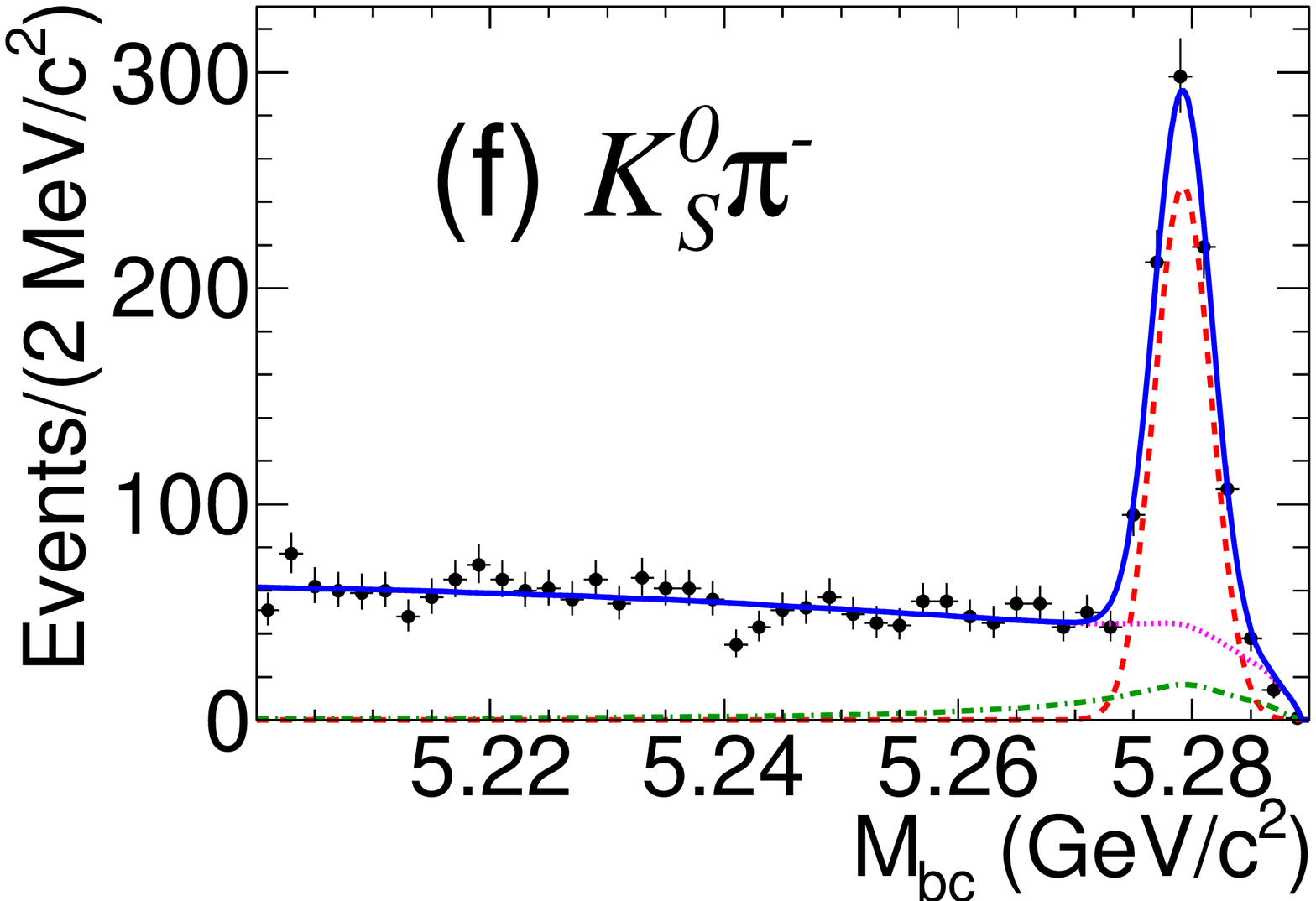} \\   
\includegraphics[width=0.3\textwidth]{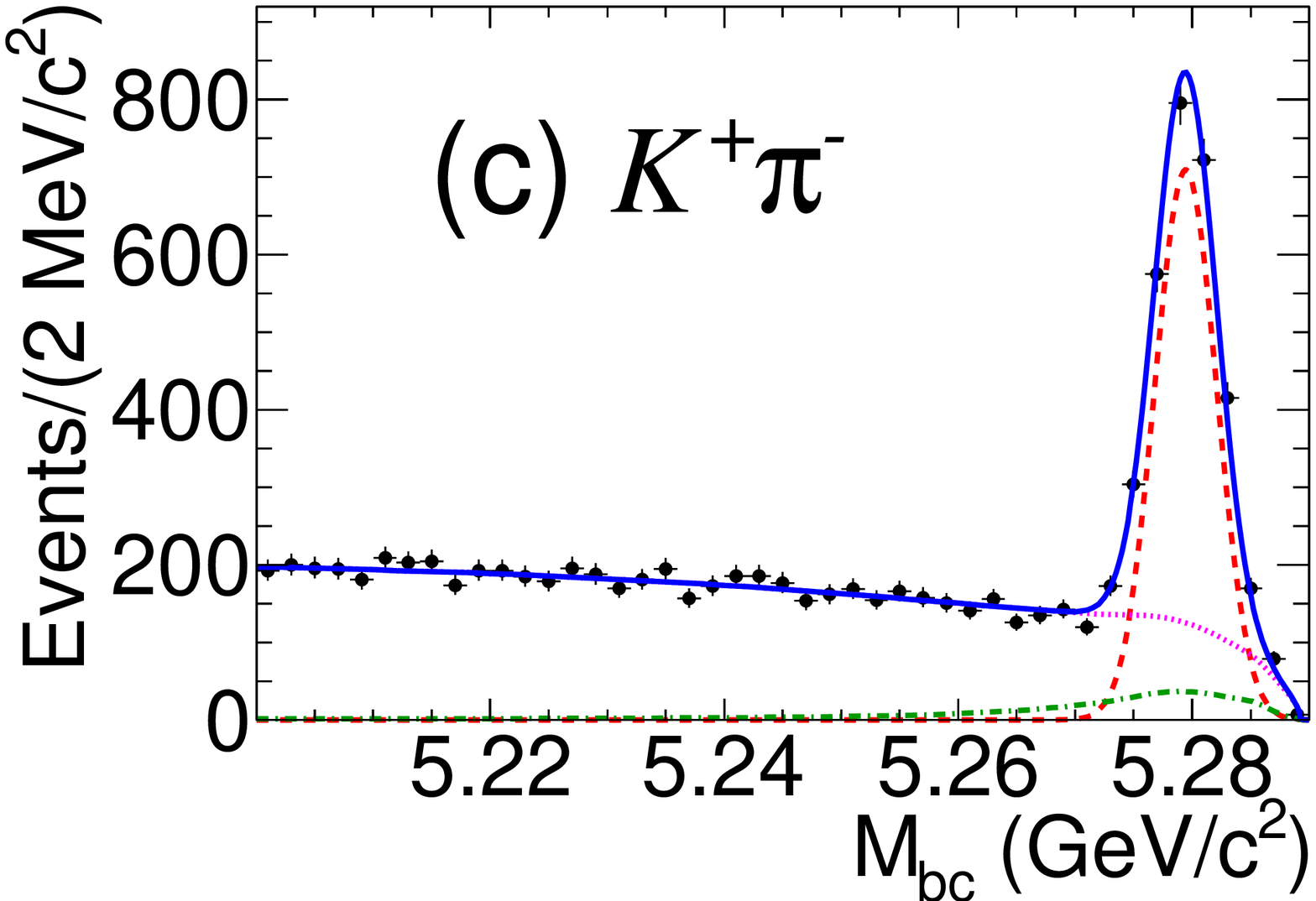} 
\includegraphics[width=0.3\textwidth]{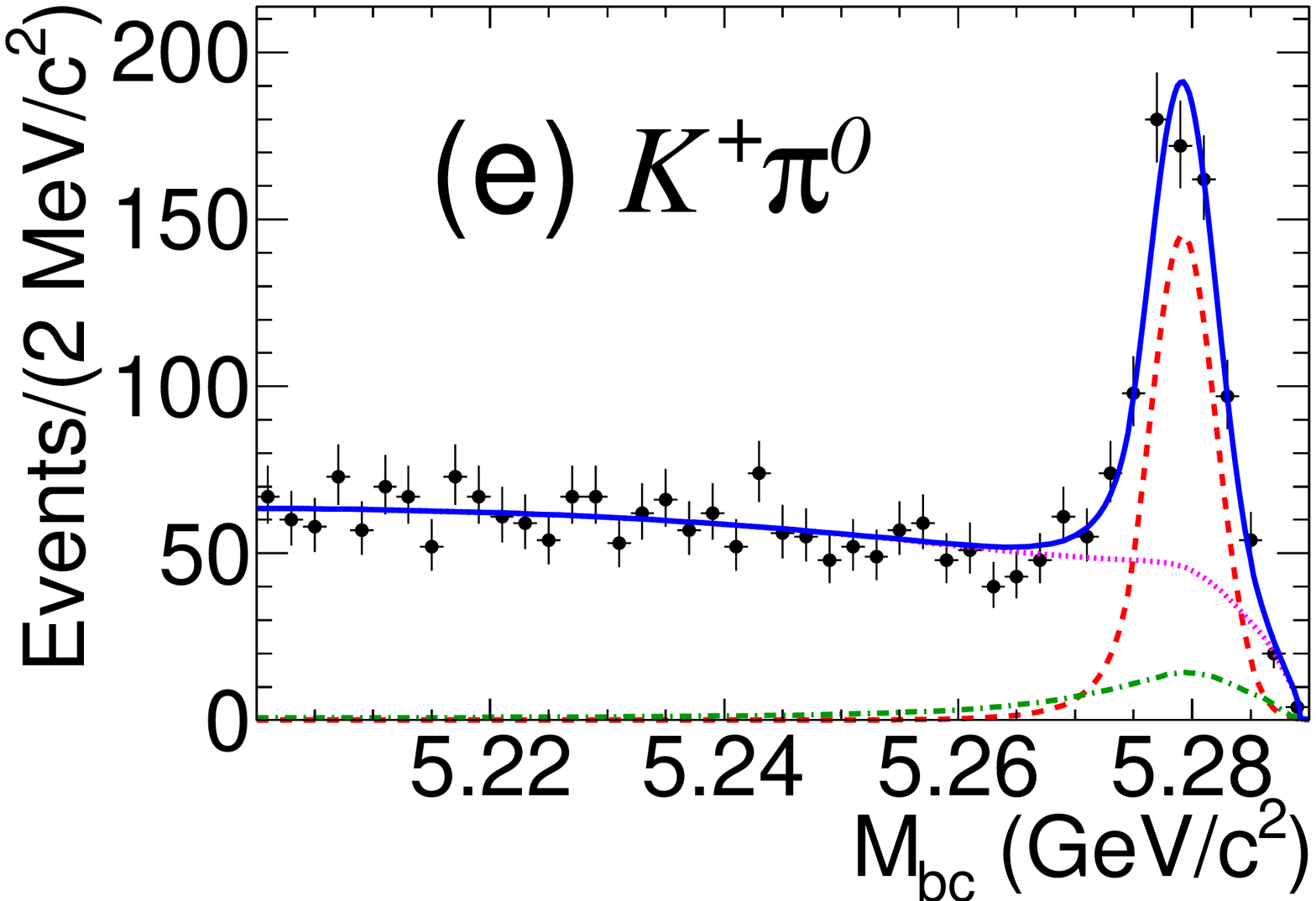} 
\includegraphics[width=0.3\textwidth]{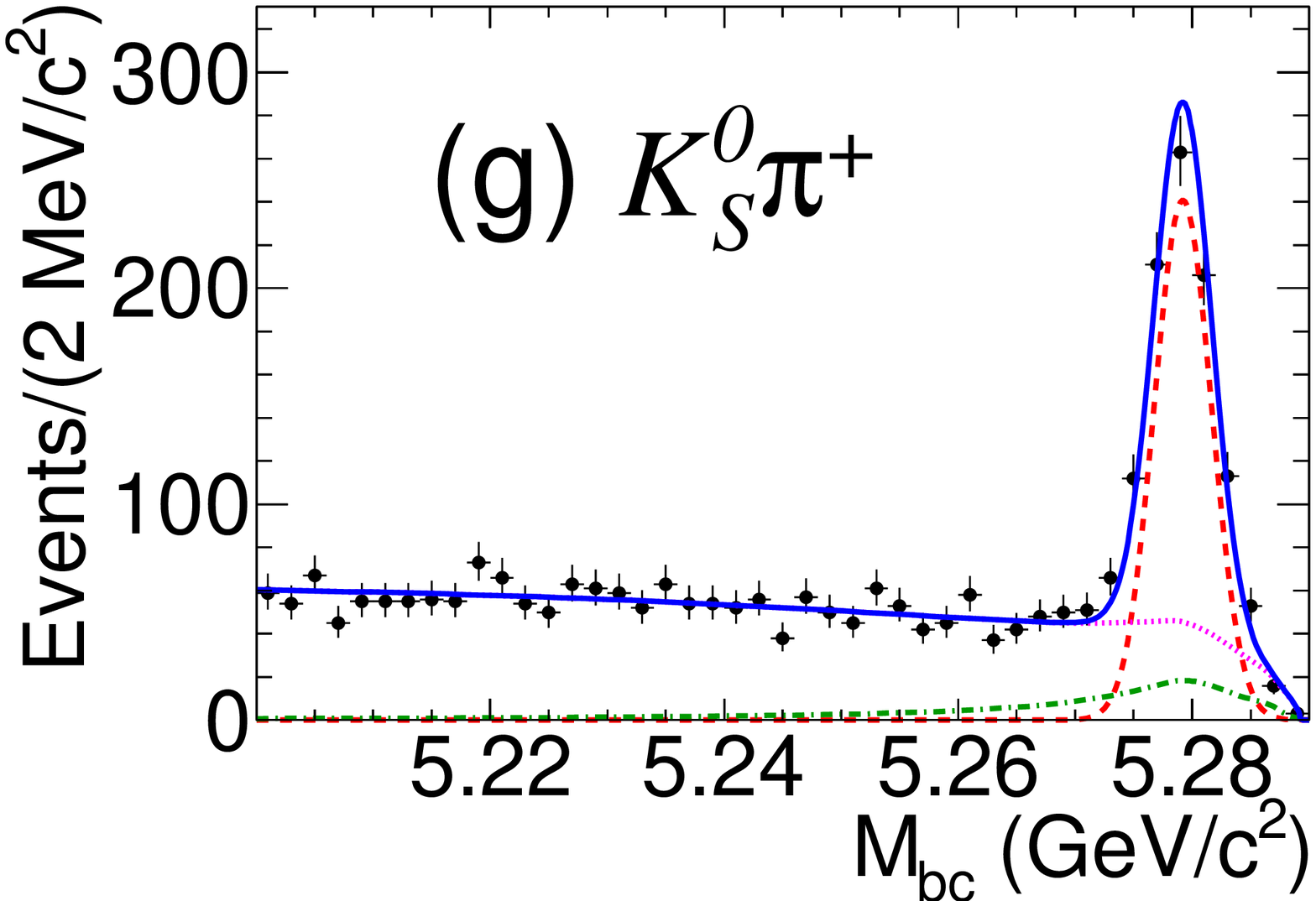}   
\caption{$\Mbc$ distributions for (a)~$K_S^0\pi^0$, (b)~$K^+\pi^-$, (c)~$K^-\pi^+$, (d)~$K^+\pi^0$ (e)~$K^-\pi^0$, (f)~$K_S^0\pi^+$, and (g)~$K_S^0\pi^-$. 
  The black points with error bars are data and solid blue curves are the total fitted distribution.
  The dashed red, dotted-dashed green, and the dotted magenta represent the signal, $B\bar{B}$ background and total background components, respectively~\cite{btosg:belle}.}
\label{fig:btosg:mbc}
\end{figure}

In the recent measurement by Belle~\cite{btosg:belle}, $B^{0}\to K^{\star 0}\gamma$ and $B^{+}\to K^{\star +}\gamma$ decays are reconstructed, where the $K^{\star}$ is formed from $K^{+}\pi^{-}$, $K_{S}^{0}\pi^{0}$,  $K^{+}\pi^{0}$ or $K_{S}^{0}\pi^{+}$ combinations (charge conjugate mode is implied unless otherwise stated).
The photon candidate is selected from isolated clusters consistent with electromagnetic shower shape.
Further, vetoes are applied on the photon candidate to ensure it does not originate from a $\pi^{0}$ or an $\eta$ decay.
The dominant background from the $e^{+}e^{-}\to q \bar{q}$ continuum events is suppressed by utilizing event shape variables in a multivariate analysis with a neural network. 
Then, a simultaneous fit is performed to all the seven $M_{\rm bc}$ distributions with the likelihood described in Ref.~\cite{btosg:belle} to extract the combined branching fractions and direct $CP$ asymmetries as well as $\Delta_{0+}$ and $\Delta A_{CP}$; the fitted distributions are shown in Figure~\ref{fig:btosg:mbc}.

The results are:
\begin{eqnarray*}
  {\cal{B}}(B^0 \to K^{*0} \gamma) &=& (3.96 \pm 0.07 \pm 0.14)\times10^{-5},\\ 
  {\cal{B}}(B^+ \to K^{*+} \gamma) &=& (3.76 \pm 0.10 \pm 0.12)\times10^{-5},\\ 
  A_{CP}(B^0 \to K^{*0} \gamma) &=& (-1.3 \pm 1.7 \pm 0.4)\%,\\ 
  A_{CP}(B^+ \to K^{*+} \gamma) &=& (+1.1 \pm 2.3 \pm 0.3)\%,\\ 
  A_{CP}(B \to K^{*} \gamma)    &=& (-0.4 \pm 1.4 \pm 0.3)\%,\\ 
  \Delta_{0+}   &=& (+6.2 \pm 1.5 \pm 0.6 \pm 1.2)\%, \\
  \Delta A_{CP} &=& (+2.4 \pm 2.8 \pm 0.5)\%,
\end{eqnarray*}

\begin{itemize}
\item Here, the first uncertainty is statistical, the second is systematic, and the third for $\Delta_{0+}$ is due to uncertainty in BF ratio for \Y4S to $B^{+}B^{-}$ and $B^{0}\Bb^{0}$.
  
\item First evidence for $\Delta_{0+}$ is found as well as first result is reported for $\Delta A_{CP}$ measurement in $B\to K^{\star}\gamma$. All the results are the most precise to date and are consistent with the previous measurements by CLEO~\cite{btosg:cleo}, Belle~\cite{btosg:belle0}, BaBar~\cite{btosg:babar}, and LHCb~\cite{btosg:lhcb}.

\item Isospin violation ($\Delta_{0+}$) is reported with a significance of $3.1\sigma$, which is consistent with the SM predictions~\cite{{btosg:th1},{btosg:th5},{btosg:th8},{btosg:th9},{btosg:th10},{btosg:th11}}. Dominant uncertainties for this measurement are statistical and due to uncertainty in BFs, \Y4S to $B^{+}B^{-}$ and $B^{0}\Bb^{0}$.

\item The $A_{CP}$ and $\Delta A_{CP}$ measurements are consistent with zero.
  
\item The BF ratio of $B^{0}\to K^{\star 0} \gamma$ and $B_{s}^{0}\to \phi \gamma$ is calculated and found to be $1.10 \pm 0.16 \pm 0.09 \pm 0.18$, which is also consistent with the previous measurement by LHCb~\cite{btosg:lhcb} as well as with the SM predictions~\cite{{btosg:th6},{btosg:th8}}.
  For this measurement only the $K^{\star 0} \to K^{+}\pi^{-}$ mode is utilized to cancel common systematic uncertainties, while the  BF for $B_{s}^{0}\to \phi \gamma$ is taken from a previous Belle measurement from 121 \invfb data recorded at the \Y5S resonance~\cite{btophig:belle}.
\end{itemize}

\section{Angular Analysis of  $B\to K^{\star}\ell\ell$}
\label{btosll}
The decay $B\to K^{\star}\ell\ell$ involves the FCNC transition $b\to s \ell \ell$, and is a rare process in the SM.
Interestingly, in the recent years several measurements have shown deviations from the SM for this decay~\cite{btosll:th1}.
A global fit to $B$ decay results suggests lepton non-universality, where muon modes would have larger contributions from the NP than electron modes~\cite{btosll:th2}. This motivates to check lepton-flavor dependent angular analysis. The observables $P_{i}'$, introduced in Ref.~\cite{{btosll:th3},{btosll:th4}}, are considered to be largely free of form-factor uncertainties~\cite{btosll:th5}. Any deviation from zero in the difference $Q_{i} = P_{i}^{\mu} - P_{i}^{e}$ would be a direct hint of NP~\cite{btosll:th6}; here, $i=4,5$ and $P_{i}^{\ell}$ refers to $P_{4,5}'$ in the corresponding lepton mode.
The definition of $P_{i}'$ values follows the LHCb convention~\cite{btosll:lhcb}.

\begin{figure}[htb]
\includegraphics[width=0.5\textwidth]{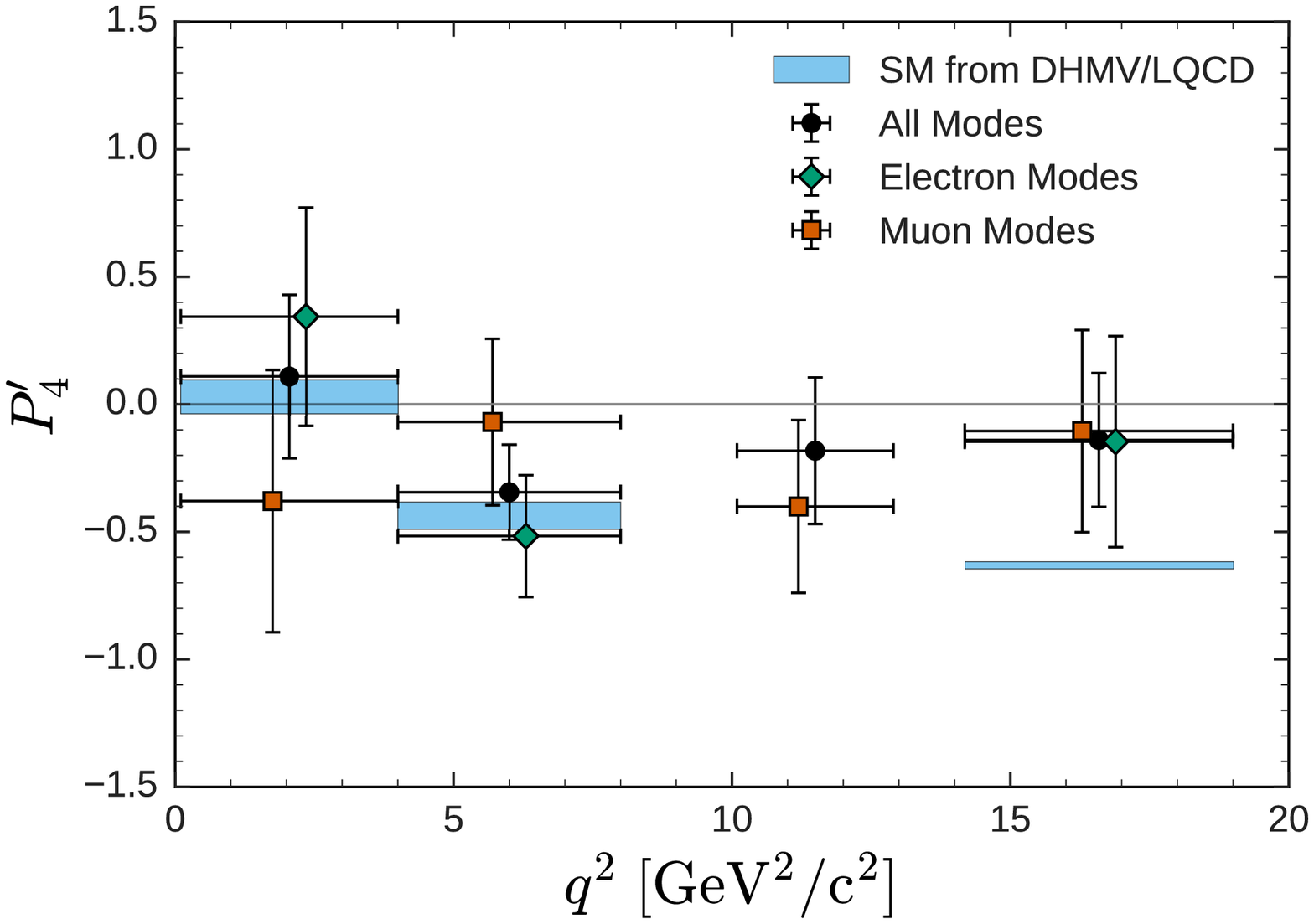}
\includegraphics[width=0.5\textwidth]{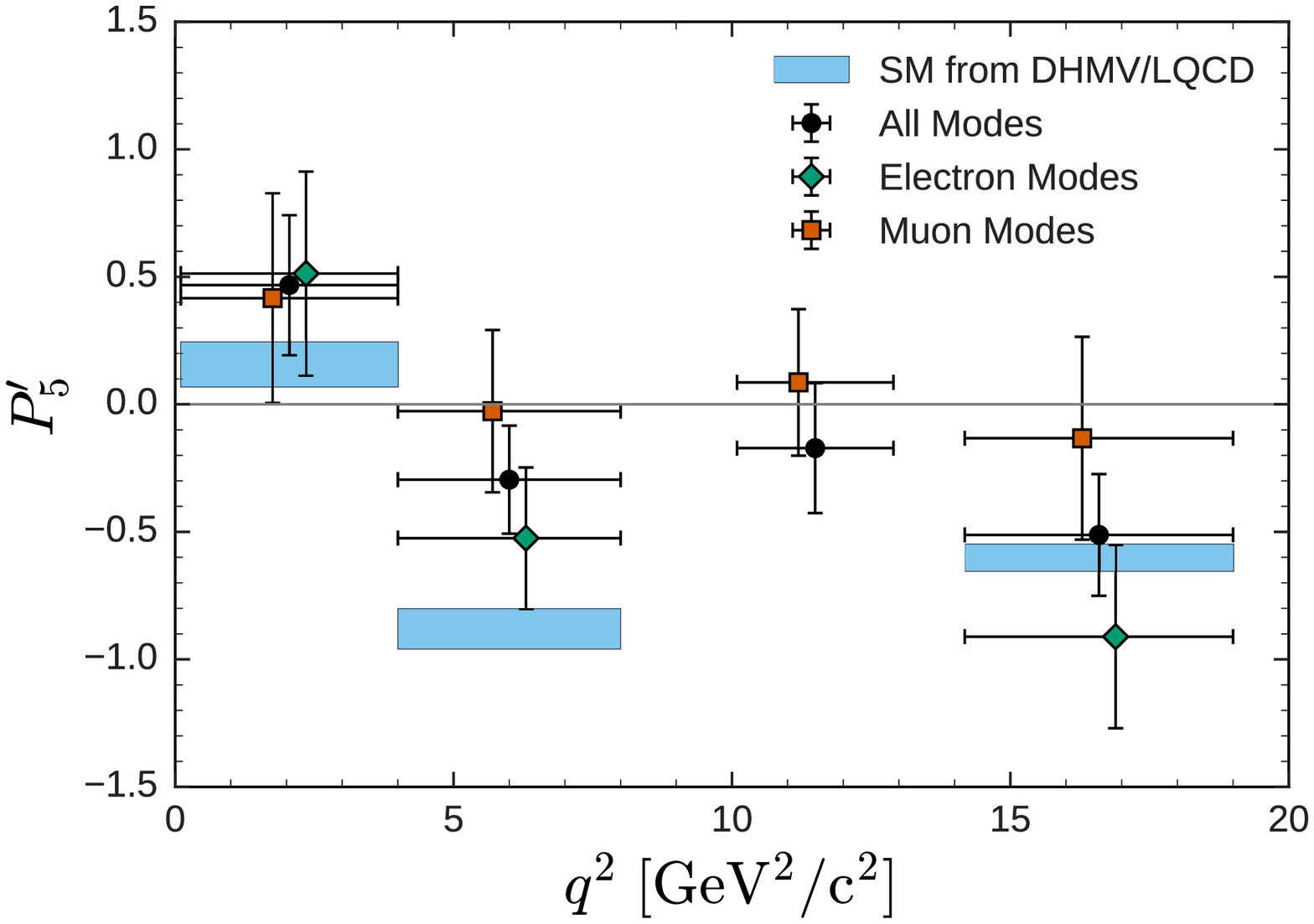}   
\caption{$P_{4}'$ (left) and $P_{5}'$ (right) observables for combined, electron and muon modes. The SM predictions are shown as cyan filled boxes~\cite{btosll:belle}.}
\label{fig:btosll:p}
\end{figure}

In total, four decay modes are reconstructed $B^{0}\to K^{\star 0} \mu^{+} \mu^{-}$,  $B^{+}\to K^{\star +} \mu^{+} \mu^{-}$, $B^{0}\to K^{\star 0} e^{+} e^{-}$ , and  $B^{+}\to K^{\star +} e^{+} e^{-}$; where $K^{\star 0}$ decays to $K^{+}\pi^{-}$ and $K^{\star +}$ decays to $K^{+}\pi^{0}$ or $K_{S}^{0}\pi^{+}$.
Signal yields are extracted with an unbinned extended maximum likelihood fit and in total, $127\pm15$ and $185\pm17$ signal candidates are obtained for the electron and muon channel, respectively.
The analysis is performed in the four independent bins of $q^{2}$ (invariant mass squared of the two leptons). An additional $q^{2}\in (1.0,6.0)$ ${\rm GeV^{2}}/c^{2}$ bin is considered, which is favored for theoretical predictions~\cite{btosll:th3}. To maximize the potency of limited statistics, a data-transformation technique is utilized~\cite{{btosll:lhcb1},{btosll:th7}}. The result is shown in Figure~\ref{fig:btosll:p}, where it is also compared with SM predictions~\cite{{btosll:th8},{btosll:th9}}.

Overall the result is in agreement with the SM value.
The largest deviation is $2.6\sigma$, observed in $q^{2}\in (4.0,8.0)$ ${\rm GeV^{2}}/c^{2}$ bin of $P_{5}'$ for the muon mode.
This tension is coincidental to the $P_{5}'$ anomaly earlier reported by LHCb~\cite{{btosll:lhcb},{btosll:lhcb1}}. In the same region the electron modes deviate by $1.3\sigma$ and the combination deviates by $2.5\sigma$. 

\begin{figure}[htb]
\includegraphics[width=0.5\textwidth]{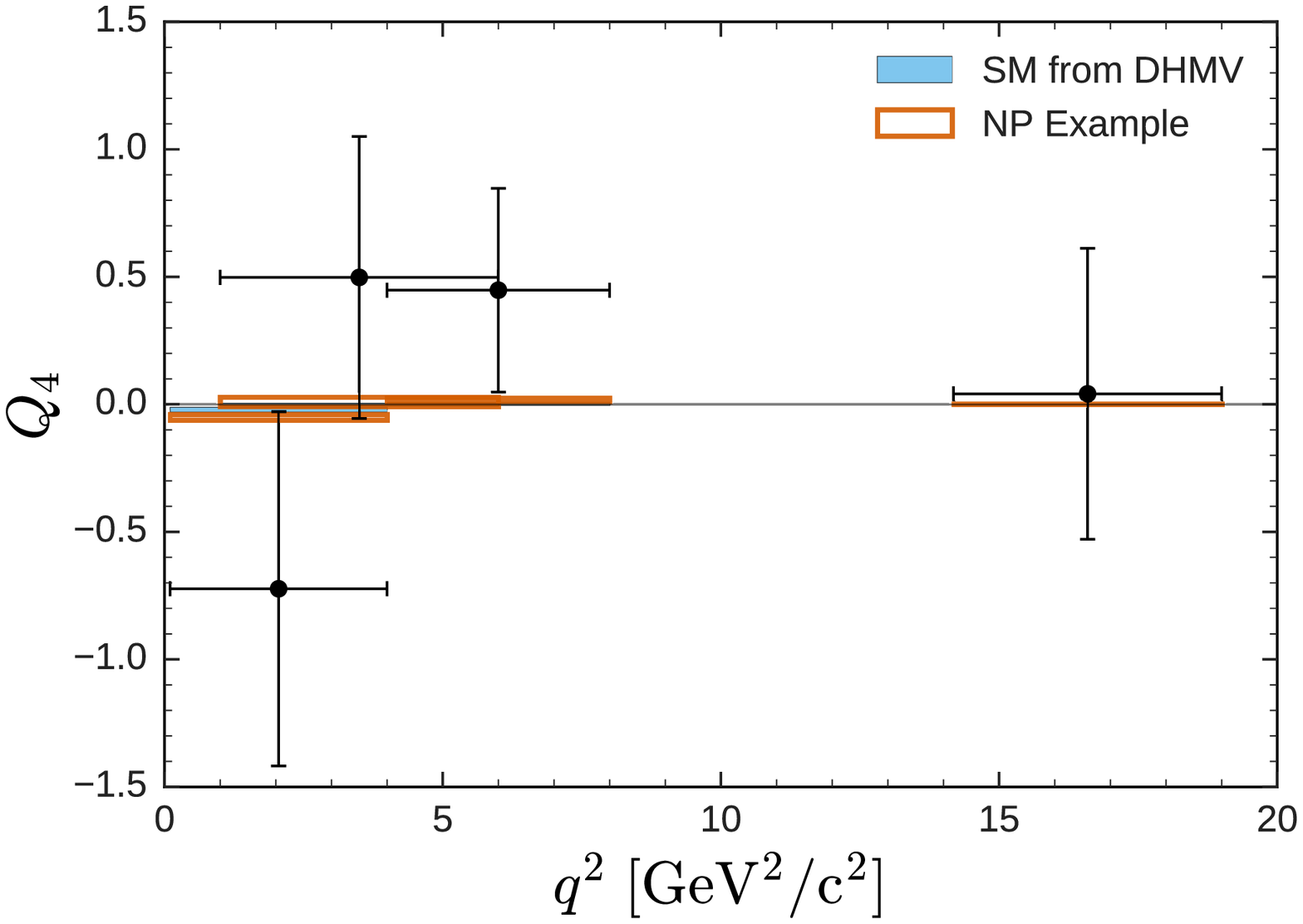}
\includegraphics[width=0.5\textwidth]{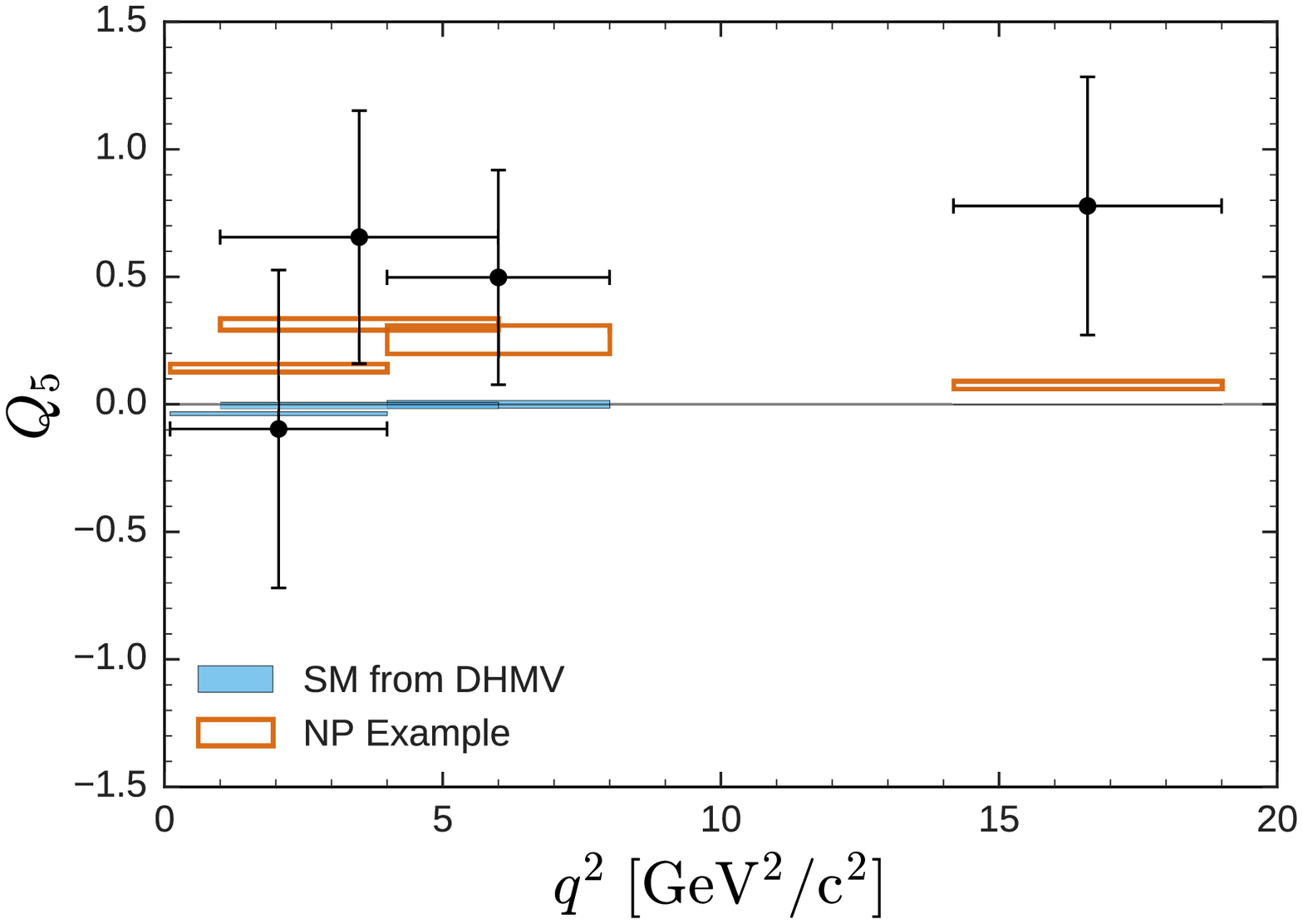}   
\caption{$Q_{4}$ (left) and $Q_{5}$ (right) observables compared with SM and NP scenario, respectively shown by the cyan filled and brown open boxes~\cite{btosll:belle}.}
\label{fig:btosll:q}
\end{figure}

The observables $Q_{4}$ and $Q_{5}$ are presented in Figure~\ref{fig:btosll:q}, where they are compared with SM and NP scenario~\cite{btosll:th6}. The results show no significant deviation from zero.

Global fits performed including these measurements~\cite{btosll:belle} suggests for lepton-universality violation~\cite{btosll:th10}.

\section{Search for $B\to h\nu\nu$}
\label{sec:btohnn}
The decays $B\to h\nu\nu$ (where $h$ refers to $K^{+}$, $K_{s}^{0}$, $K^{\star +}$, $K^{\star 0}$, $\pi^{+}$, $\pi^{0}$, $\rho^{+}$ or $\rho^{0}$) are theoretically clean due to the exchange of a $Z$ boson alone, in comparison to other $b \to s$ transitions where the virtual photon also contributes~\cite{btohnn:th1}.

\begin{figure}[htb]
  \centering
  \includegraphics[width=0.75\textwidth]{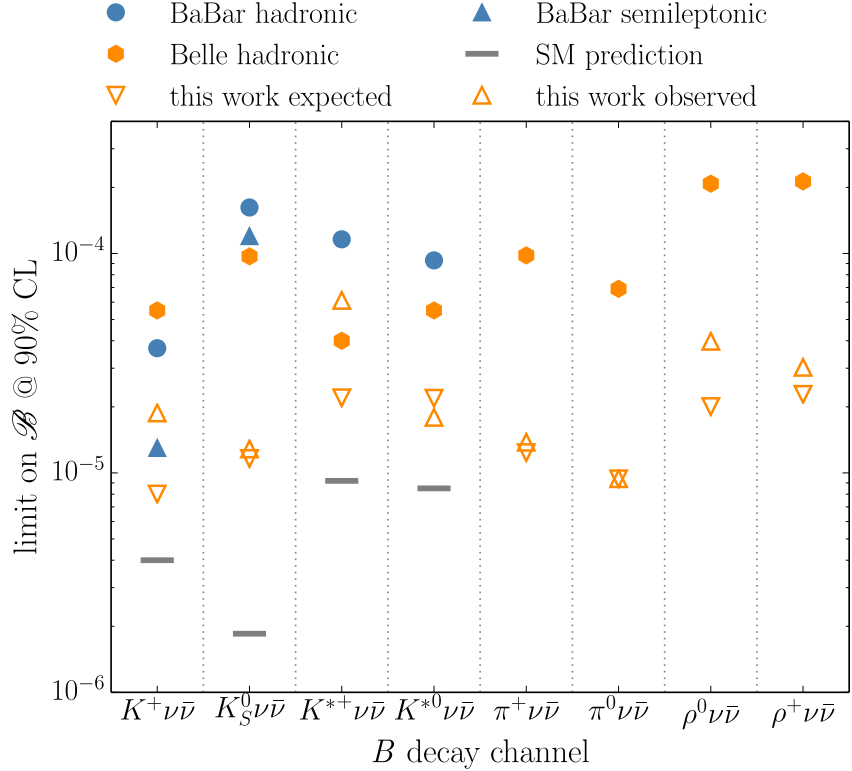}   
  \caption{Observed upper limits along-with the expected values and previous measurement. SM predictions are also shown for the $K^{(\star)}$ modes~\cite{btohnn:belle}.}
  \label{fig:btohnn}
\end{figure}

Previously, the decays $B\to h\nu\nu$ have been searched in Belle utilizing the  hadronic tag method~\cite{btohnn:belle0} and in BaBar using both hadronic~\cite{btohnn:babar1} and semi-leptonic tag~\cite{btohnn:babar2}.
The recent Belle analysis~\cite{btohnn:belle} is based on a more efficient semi-leptonic tagging method.
The signal $B$ daughter candidates are reconstructed through the decays: $K^{*0}\to K^{+}\pi^{-}$, $K^{\star +}\to K^{+}\pi^{0}$ and $K_{s}^{0}\pi^{+}$, $\rho^{+}\to\pi^{+}\pi^{0}$, $\rho^{0}\to \pi^{+}\pi^{-}$, $K_{s}^{0}\to \pi^{+}\pi^{-}$, and $\pi^{0}\to \gamma\gamma$. Event shape variables are  utilized to suppress continuum events. Signal events are identified from extra energy in the electromagnetic calorimeter ($E_{ECL}$), which is calculated by removing all the associated $ECL$ energy from tag and signal $B$ mesons. T
he largest signal contribution is observed in the decay $B\to K^{\star +}\nu\nu$ with a significance of $2.3\sigma$.
In the absence of a significant signal in any of the modes, upper limits on the BFs are measured with 90\% confidence level. The result is shown in Figure~\ref{fig:btohnn} along with expected values and previous measurements.

These decays can be observed with Belle II~\cite{belle2:tdr}, assuming the SM prediction holds.
Belle II will be able to provide a measurement with uncertainties of similar size as the current theoretical uncertainties~\cite{belle2}.

\section{Summary}
\label{sec:sum}
FCNC processes like $b\to s$ transitions are forbidden at tree level in the SM, however various NP contributions can interfere with these types of rare SM amplitudes.
Recently, the first evidence of isospin violation is reported in the $B\to K^{\star} \gamma$ decay~\cite{btosg:belle}; also first measurement of the difference of $CP$ asymmetries, between charged and neutral $B$ meson in performed in the same analysis.
Belle reports the most precise measurement of the BFs, direct $CP$, and isospin asymmetries, and the results are consistent with SM predictions and previous measurements~\cite{{btosg:th2},{btosg:th5},{btosg:th7},{btosg:th8},{btosg:th9},{btosg:th10}}.
First lepton-flavor dependent angular analysis for the decay $B\to K^{\star}\ell\ell$ is performed~\cite{btosll:belle}; and results are consistent with both SM values and NP scenarios.
Belle set the most stringent upper limits on the BFs for the decay $B\to h\nu\nu$~\cite{btohnn:belle}.
The upper limits are close to SM predictions~\cite{btohnn:th1} for the $K^{(\star)}$ modes and Belle II has brighter prospects to observe these decays.

\end{document}

%% file: econfmacros.tex
\def\beq{\begin{equation}}
\def\eeq#1{\label{#1}\end{equation}}
\def\eeqn{\end{equation}}

\def\beqa{\begin{eqnarray}}
\def\eeqa#1{\label{#1}\end{eqnarray}}
\def\eeqan{\end{eqnarray}}

\let\bar=\overbar

\def\etal{{\it et al.}~}

\def\Dslash{\not{\hbox{\kern-4pt $D$}}}
\def\dslash{\not{\hbox{\kern-2pt $\del$}}}

\def\msb{{\bar{\ssstyle M \kern -1pt S}}}

\def\Bbar   {\kern 0.18em\overline{\kern -0.18em B}{}\xspace}
\def\Bb     {\ensuremath{\Bbar}\xspace}
\def\BB     {\ensuremath{B\Bbar}\xspace}

\def\Kbar   {\kern 0.18em\overline{\kern -0.18em K}{}\xspace}
\def\Kb     {\ensuremath{\Kbar}\xspace}
\def\Mbc    {M_{\rm bc}}

